# Electron spin relaxation in *X*-valley of indirect bandgap Al$_x$Ga$_{1-x}$As: A new horizon for the realization of next generation spin-photonic devices

Priyabrata Mudi,[1,3,a)] Shailesh K. Khamari,[1] Joydipto Bhattacharya,[2,3] Aparna Chakrabarti,[2,3] and Tarun K. Sharma[1,3,b)]

[1]Semiconductor Materials Lab, Materials Science Section, Raja Ramanna Centre for Advanced Technology, Indore, 452013, Madhya Pradesh, India

[2]Theory and Simulations Laboratory, HRDS, Raja Ramanna Centre for Advanced Technology, Indore - 452013, 452013, Madhya Pradesh, India

[3]Homi Bhabha National Institute, Training School Complex, Anushakti Nagar, Mumbai, 400094, Maharashtra, India

GaAs/AlGaAs quantum well (QW) system is utilized to investigate the electron spin relaxation in the satellite *X-valley* of indirect band gap Al$_{0.63}$Ga$_{0.37}$As epitaxial layers through polarization resolved photo-luminescence excitation spectroscopy. Solving the rate equations, steady state electronic distribution in various valleys of Al$_x$Ga$_{1-x}$As is estimated against continues photo carrier generation and an expression for the degree of circular polarization (DCP) of photoluminescence coming from the adjacent quantum well (QW) is derived. Amalgamating the experimental results with analytical expressions, the *X-valley* electron spin relaxation time ($\tau_S^X$) is determined to be 2.7 ± 0.1 ps at 10 K. To crosscheck its validity, theoretical calculations are performed based on Density Functional Theory within the framework of the projector augmented wave method, which support the experimental result quite well. Further, temperature dependence of $\tau_S^X$ is studied over 10-80 K range, which is explained by considering the intra-valley scattering of carriers in the *X-valley* of indirect band gap AlGaAs barrier layer. It is learnt that the strain induced modification of band structure lifts the degeneracy in *X-valley*, which dominates the electron spin relaxation beyond 50 K. Furthermore, the DCP spectra of hot electrons in indirect band gap AlGaAs layers is found to be significantly different compared to that of direct bandgap AlGaAs. It is understood as a consequence of linear *k* dependent Dresselhaus spin splitting and faster energy relaxation procedure in the *X-valley*. Findings of this work could provide a new horizon for the realization of next generation spin-photonic devices which are less sensitive to Joule heating.

# I. Introduction

For a significant amount of time, the Spin dynamics of photo-generated electrons in III-V semiconductors has gained substantial curiosity of researchers due to its applications in the development of fast and high performance spin-optoelectronic devices.[1-5] Following the works of Dyakonov and Perel in the early 70's, researchers have developed an understanding of the electron spin dynamics in III-V semiconductors both theoretically as well as experimentally.[6-12] The theoretical substructure of spin optoelectronics is developed based on Density Functional Theory and 14-34 band *k.p* theory.[6, 9-10] On the other hand, the experimental development of electron spin dynamics is extensively established through polarization and time resolved photoluminescence (PL) spectroscopy, inverse spin Hall effect and Kerr rotation microscopy.[13-15] Interestingly, most of these works are restricted to the near band edge region of direct band gap III-V semiconductors and the satellite valleys are vastly ignored. However, for practical implications of spintronic devices, like spin light emitting devices (LED) and spin lasers, upper valleys play a pivotal role since these devices operate under high bias conditions where electrons in general occupy all the three valleys ($\Gamma$, $L$ and $X$).[16, 17] Further, an immense spin-orbit coupling is also predicted in the satellite valleys as *p* and *d* type orbitals from the higher bands overlap with the conduction band orbital wave function at higher *k* values and are expected to influence its nature.[18] Based on this, Okamoto et. al[19] explored the *L-valley* electrons spin dynamics in n-GaAs through Inverse spin Hall effect by predicting a huge spin Hall angle. Furthermore, linear *k*-dependent spin splitting and high Lande *g* factor of electrons are the two important features of satellite valleys that could help in improving the performance of spintronic devices. Due to these reasons, spintronics devices fabricated based on the satellite valleys of III-V semiconductors have caught the attention of the research community in recent times.[16, 20-22] Nonetheless, Joule heating in the device due to application of high bias is one of the major drawbacks in their practical implication.[20] Further, a steep fall of spin relaxation time is seen under high bias[19] as electrons are transferred to the satellite *L* valleys where the reported value of spin relaxation time ($\tau_S^L$) is very fast.[22] Recently, it was discovered that one can optically inject spin polarized electrons in the satellite *L-valleys* of GaAs and AlGaAs.[22,23] This could become a powerful tool in overcoming the Joule heating problem since the satellite valleys are now optically accessible. Amid the importance of optical spin injection in the satellite valleys, this technique is seldom discussed in literature. The main reason behind this lies in the direct band gap nature of most of III-V semiconductors where only a very small fraction of photo-generated carriers occupies the higher satellite valleys. In this regard, indirect bandgap semiconductors like $Al_xGa_{1-x}As$ (x > 0.4) provide an excellent opportunity since their satellite *X-valley* minimum lie at lower energy compared to *Γ-valley* minima. Therefore, in spite of photo-injection in *Γ*-valley, most of the electrons thermalize to satellite valleys within tens of femtosecond[24] even in absence of applied bias. Further, AlGaAs is almost lattice matched to GaAs and high quality epitaxial layers can be easily grown on GaAs wafers by growth techniques like Metal Organic Vapour Phase Epitaxy (MOVPE) and Molecular Beam Epitaxy (MBE). Nevertheless, fabrication of spin-photonic devices requires a complete knowledge of optical spin injection process, spin dynamics and spin relaxation time in that particular material. Thus, a knowledge of the spin relaxation time of electrons in satellite valleys is extremely important in order to optimize the performance of associated devices. Zhang et al.[22] have determined the electron spin relaxation time in the *L*-valley of GaAs through time and polarization resolve photoluminescence excitation (PLE) spectroscopy and its value is estimated to be as short as 0.2 ps. However, no such information is yet available in literature for *X-valley*, which is the lower most valley for indirect band gap AlGaAs. Thus, optical spin orientation spectra as well as the spin relaxation time of electrons in the *X-valley* of indirect AlGaAs needs to be evaluated before its Spin-photonic applications. As pointed earlier, in case of direct bandgap semiconductors, optical orientation of electrons is investigated through polarization resolved PLE spectroscopy by measuring the degree of circular polarization (DCP) of photoluminescence (PL) as a function of excitation energy. Though this technique is non-destructive in nature, it is irrelevant for indirect bandgap semiconductors due to the suppression of radiative processes that gives rise to a weak PL signal at low temperature. Even though, the magnitude of indirect band gap PL increases with temperature, any possible advantage because of this is shadowed by a simultaneous reduction of spin relaxation time.[13] Thus, a high intensity tuneable laser source may be required to increase the PL magnitude, which will allow the measurement of a few percent of DCP of the signal. However, use of high intensity laser sources to increase the PL intensity will not be of much help as electron spin dynamics itself depends on the intensity of excitation source.[14] In this context, GaAs/AlGaAs quantum wells (QW) offers a credible system to investigate the optical orientation properties of electrons in the barrier layer using low intensity light sources. In our previous work [23] we have employed GaAs/AlGaAs multi QW architecture to study the optical injection of spin polarized electrons in $Al_{0.22}Ga_{0.78}As$ material over the excitation energy range of 1.85 – 3.4 eV using a Xenon arc lamp (average power density ~ 4 mW/cm$^2$). Based on the results, it was established that although the magnitude of DCP is determined by QW parameters, its spectral dependence is solely governed by the barrier layer and is identical to the bulk material. In this article, we extend the method to indirect bandgap materials since the radiative recombination of carriers can occur efficiently in QW layer. Spin relaxation of photo injected electrons in *X-valley* of indirect bandgap AlGaAs epilayers is successfully measured in 10-80 K range. Theoretical framework along with associated experiments are discussed with an aim of the measurement of $\tau_S^X$, which is found to be an order of magnitude higher than $\tau_S^L$, thus making *X-valley* of AlGaAs to be a new arena for the development of spin-photonic devices.

The rest of the article is organized as follows. In Section II, a theoretical model based on the capture of electrons in QW via different valleys, *Γ* to QW in direct and *Γ-X* to QW in indirect AlGaAs barrier layer, is proposed. The corresponding rate equations are solved to estimate the steady state spin polarized carrier density in barrier and QW layers. Timescales of different phenomena like inter-valley scattering, spin relaxation, quantum capture and carrier recombination are included in the rate equations in order to explain the experimental observations. Section III contains the sample details and experimental methods that are used for characterization. Section IV consists of miscellaneous experimental results, theoretical calculations based on the Density Functional Theory and comparison of the estimated value of *X-valley* electrons spin relaxation time. Analysis of excitation energy and temperature dependent electron spin dynamics is also included in this section. Further, section V brings forth a brief summary of the work presented in the article.

## II. Theoretical modelling

The rate equation model developed by Stanton et al.[24] and then extended by Mudi et al.[25] estimates the steady state distribution of photo-excited spin polarized electrons in different valleys of conduction band, by considering different scattering and relaxation mechanism viz. spin and energy relaxation, carrier capture in QW and carrier recombination. Here, we have further extended this model, to obtain a steady state expression for the DCP of PL signal originating from the QW layer. Note that the model of Mudi et al.[25] excludes the dynamics of photo generated holes due to their ultrafast spin relaxation time.[9] In this context, it is necessary to evaluate the photo generated spin polarized electron population ($n_S^{QW}$) as well as total electron population ($n_{Total}^{QW}$) in the ground state of QW layer for the cases of photo excitation in barrier and QW layers. One can derive an expression for DCP from the ratio of these two quantitates.

Since we are interested in the electron spin dynamics in the barrier material, the contribution arising from the QW must be ruled out. This is why the rate equations are solved for photo excitation in the barrier as well as in the QW and the contribution from QW are carefully separated out. Further, Pfalz et al. [26] have derived an expression to understand how initial spin relaxation during energy relaxation from the higher states of QW affects the measured value of DCP. According to their calculations, the fraction of electrons ($R$) which will be able to retain their initial spin polarization after reaching the QW ground state is given by the formula, [26]

$$R = 1 - \left(\frac{2\pi\tau_E}{T_Z}\right)^2 \tag{1}$$

where, $\tau_E$ is the energy relaxation time within the QW, which is of the order of hundred femto-second[26] and $T_Z$ is the average spin relaxation time of electrons in the QW excited states. [26] For thick QWs, $T_Z$ is estimated to be less sensitive to electrons kinetic energy ($E_K$) as compared to thin QWs. [26] For a 15 nm thick QW, calculated value of $T_Z$ is approximately 10 ps for $E_K = 75$ meV. These parameters yield the numerical value of $R$ to be 0.996. From this, one can conclude that for thick QWs, photo-generated electrons preserve significant spin polarization while energy relaxation and two level rate equation will be sufficient to understand their dynamics, the two levels being the ground state of barrier and QW. In this context, the ascribed rate equation can be written as,

$$\frac{d}{dt}\begin{bmatrix} n_S^B \\ n_S^{QW} \end{bmatrix} = \begin{bmatrix} -(\gamma_C + \gamma_S^B) & 0 \\ \gamma_C & -\gamma_S^{QW} \end{bmatrix}\begin{bmatrix} n_S^B \\ n_S^{QW} \end{bmatrix} + \begin{bmatrix} GP_0 \\ 0 \end{bmatrix} \tag{2}$$

where, $n_S^B$ is the steady state photo generated spin polarized electron population in the barrier layer, $\gamma_C$, and $\gamma_S^B$ ($\gamma_S^{QW}$) stand for the electron capture rate from barrier to QW ground state, and electron spin relaxation rate in barrier layer (QW ground state) respectively. Further, $G$ and $P_0$ are the photo generation rate and the instantaneous degree of spin polarization of photo generated electrons in the barrier layer respectively. Note that $\gamma_S^{QW}$ also incorporates the loss of electron spin due to the electron-hole recombination in QW. More precise expression for $\gamma_S^{QW}$ would be $\gamma_S^{QW} = \gamma_S^{pure} + \gamma_R^{QW}$. Here, $\gamma_S^{pure}$ and $\gamma_R^{QW}$ are the pure spin relaxation rate and electron recombination rate in the quantum well ground state. Now, from the steady state solution of Eq. (2), one can easily acquire,

$$\begin{bmatrix} n_S^B \\ n_S^{QW} \end{bmatrix} = G \begin{bmatrix} \dfrac{P_0}{(\gamma_C + \gamma_S^B)} \\ \dfrac{P_0 \gamma_C}{\gamma_S^{QW}(\gamma_C + \gamma_S^B)} \end{bmatrix} \quad (3)$$

Further, in order to derive an expression for $n_{Total}^{QW}$, all spin relaxation channels should be put to zero and $P_0$ should be equal to 1. This gives rise to the expression,

$$\begin{bmatrix} n_{Total}^B \\ n_{Total}^{QW} \end{bmatrix} = \begin{bmatrix} \dfrac{G}{\gamma_C} \\ \dfrac{G}{\gamma_R^{QW}} \end{bmatrix} \quad (4)$$

Thus, the formula to evaluate the DCP of PL signal coming from the QW for photo excitation in the barrier is written as,

$$DCP = \frac{n_S^{QW}}{n_{Total}^{QW}} = \frac{P_0}{\left(1+\dfrac{\gamma_S^{pure}}{\gamma_R^{QW}}\right)\left(1+\dfrac{\gamma_S^B}{\gamma_C}\right)} \quad (5)$$

This expression is quite similar to the DCP expression obtained by other researchers for a two level system, [22] indicating the accuracy of the rate equation model presented here.

### (a). Direct band gap AlGaAs barrier

A pictorial representation of electron spin dynamics for direct band gap AlGaAs barrier is shown in Fig. 1 (a). In this case, electrons are photo generated in the $\Gamma$-valley of AlGaAs barrier which are directly captured by the quantum well after diffusion and subsequent quantum capture via interaction with optical phonons. Thus, the expression of DC$P$ for photo excitation in the barrier layer can be written as,

$$DCP^\Gamma = \frac{n_S^{QW}}{n_{Total}^{QW}} = \frac{P_0}{\left(1+\dfrac{\gamma_S^{pure}}{\gamma_R^{QW}}\right)\left(1+\dfrac{\gamma_S^\Gamma}{\gamma_C}\right)} \quad (6)$$

Here, $\gamma_S^B$ is replaced with $\gamma_S^\Gamma$ as the electron spin dynamics is limited in the $\Gamma$-valley of AlGaAs barrier. The superscript $\Gamma$ with DCP term is defined to highlight that the carriers are being captured from the $\Gamma$ valley of AlGaAs barrier.

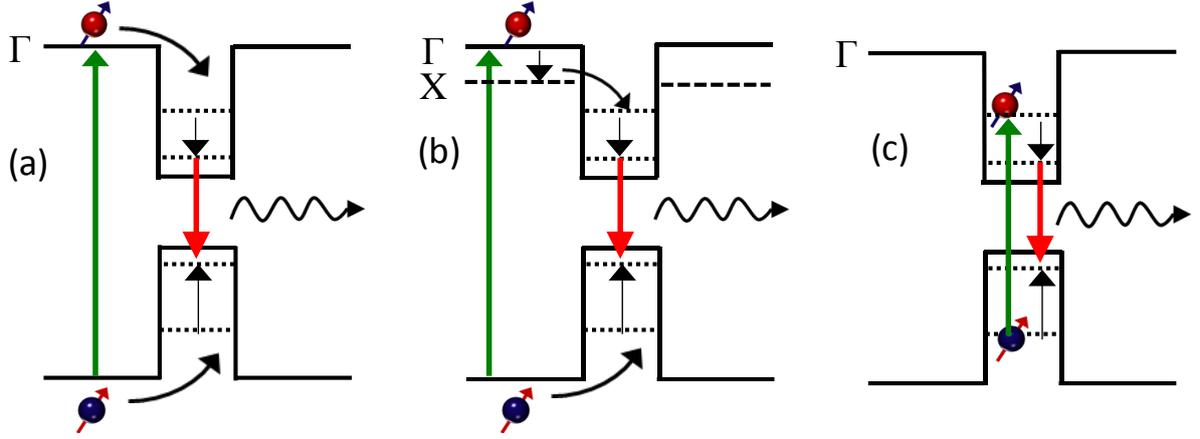

*Fig. 1. Schematic diagram of electron capture and recombination in AlGaAs/GaAs Quantum Well (QW) for photo excitation in the (a) direct band gap barrier (b) indirect band gap barrier, (c) QW.*

**(b). Indirect band gap AlGaAs barrier**

For indirect band gap AlGaAs barrier electron spin dynamics is slightly different. Here, electrons are photo excited in the $\Gamma$ valley, which lies above the satellite $L$ and $X$ valleys in the *E-k* diagram. Since in indirect band gap AlGaAs, the conduction band minimum lies near the $X$ point, photo excited electrons will eventually thermalize to *X-valley* via energy relaxation process, which makes it a multi-level system. Therefore, writing a closed form equation like Eq. (5) is not straight forward. In this case, it is necessary to determine steady state spin polarized carrier population in $\Gamma$, $L$, and $X$ valleys prior to being captured by the QW. Note that the thermalization procedure here might occur via a single step process ($\Gamma$ to $X$) or dual step ($\Gamma$ to $L$ and then $L$ to $X$) process. The ascribed rate equations are,

$$\frac{d}{dt}\begin{bmatrix} n_S^\Gamma \\ n_S^L \\ n_S^X \end{bmatrix} = \begin{bmatrix} -(\gamma^{\Gamma-L}+\gamma^{\Gamma-X}+\gamma_S^\Gamma) & 0 & 0 \\ \gamma^{\Gamma-L} & -(\gamma^{L-X}+\gamma_S^L) & 0 \\ \gamma^{\Gamma-X} & \gamma^{L-X} & -\gamma_S^X \end{bmatrix} \times \begin{bmatrix} n_S^\Gamma \\ n_S^L \\ n_S^X \end{bmatrix} + \begin{bmatrix} GP_0 \\ 0 \\ 0 \end{bmatrix} \quad (7)$$

where, $n_S^I$ and $\gamma_S^I$ are the density of the spin polarized carriers and spin relaxation rate in the $I^{th}$ valley respectively and $\gamma^{I-J}$ stands for the inter-valley scattering rate from $I^{th}$ to $J^{th}$ valley. Other symbols have their usual meaning. By solving these rate equations, the steady state carrier density in different valleys ($n_S^\Gamma, n_S^L, n_S^X$) can be estimated. Further, the fractional population of the spin polarized carriers can be realized by the expression,

$$\begin{bmatrix} f_S^\Gamma \\ f_S^L \\ f_S^X \end{bmatrix} = \begin{bmatrix} \dfrac{n_S^\Gamma}{(n_S^\Gamma + n_S^L + n_S^X)} \\ \dfrac{n_S^L}{(n_S^\Gamma + n_S^L + n_S^X)} \\ \dfrac{n_S^X}{(n_S^\Gamma + n_S^L + n_S^X)} \end{bmatrix} = \begin{bmatrix} \dfrac{\gamma_S^X (\gamma^{L-X} + \gamma_S^L)}{(\gamma^{L-X}\gamma_S^X + \gamma_S^X \gamma_S^L + \gamma^{\Gamma-L}\gamma_S^X + \gamma^{L-X}\gamma^{\Gamma-L} + \gamma^{\Gamma-X}\gamma^{L-X} + \gamma^{\Gamma-X}\gamma_S^L)} \\ \dfrac{\gamma^{\Gamma-L}\gamma_S^X}{(\gamma^{L-X}\gamma_S^X + \gamma_S^X \gamma_S^L + \gamma^{\Gamma-L}\gamma_S^X + \gamma^{L-X}\gamma^{\Gamma-L} + \gamma^{\Gamma-X}\gamma^{L-X} + \gamma^{\Gamma-X}\gamma_S^L)} \\ \dfrac{(\gamma^{L-X}\gamma^{\Gamma-L} + \gamma^{\Gamma-X}\gamma^{L-X} + \gamma^{\Gamma-X}\gamma_S^L)}{(\gamma^{L-X}\gamma_S^X + \gamma_S^X \gamma_S^L + \gamma^{\Gamma-L}\gamma_S^X + \gamma^{L-X}\gamma^{\Gamma-L} + \gamma^{\Gamma-X}\gamma^{L-X} + \gamma^{\Gamma-X}\gamma_S^L)} \end{bmatrix} \quad (8)$$

Here $f_S^I$ indicates the fractional population of spin polarized carriers in the $I^{th}$ valley. Finally, by putting $\gamma_S^X = 4 \times 10^{11}\ s^{-1}$ (Theoretically calculated value explained in next section), $\gamma_S^L = 5 \times 10^{12}\ s^{-1}$ (Ref. 22), $\gamma^{\Gamma-X} = 2 \times 10^{13}\ s^{-1}$ (Ref. 27), $\gamma^{\Gamma-L} = 8.3 \times 10^{12}\ s^{-1}$ (Ref. 27) and $\gamma^{L-X} = 4 \times 10^{12}\ s^{-1}$ (Ref. 28), the numerical values of $f_S^\Gamma$, $f_S^L$ and $f_S^X$ turn out to be 0.015, 0.015 and 0.97 respectively. These results are a consequence of faster $\Gamma$–$X$ inter-valley scattering time whose value is about 50 femto-seconds (*fs*) for the current scenario (Ref. 27). As a result, both $\Gamma$ and $L$ valley of indirect band gap AlGaAs barrier will play minimal role in the attribution of electron spin dynamics and majority of electrons will not lose their spin information before reaching $X$ valley. A similar concept is found in case of indirect band gap Germanium where electrons are scattered from $\Gamma$ valley to $L$ valley. [29] Therefore, it is safe to assume that photo generation of spin polarized carriers is taking place only in satellite *X-valley* and $\Gamma$ and $L$ valley have very minimal role to play. Thus, the electron spin dynamics can again be evaluated by a two state model as depicted in Fig. 1. (b). In this context, the expression for DCP of PL is given by,

$$DCP^X = \frac{n_S^{QW}}{n_{Total}^{QW}} = \frac{P_0}{\left(1 + \dfrac{\gamma_S^{pure}}{\gamma_R^{QW}}\right)\left(1 + \dfrac{\gamma_S^X}{\gamma_C^X}\right)} \quad (9)$$

where symbols have their usual meaning and the superscripts with the symbols identify the region of occupation of photo generated electrons. This expression is very similar to the expression of DCP derived for direct band gap barriers [Eq. (6)], except that $\gamma_S^\Gamma$ is replaced by $\gamma_S^X$, representing the different capture path of the photo generated electrons in the barrier, while the contribution coming from the QW remains the same. Further, it should be emphasized that, in the present model, $X$-$\Gamma$ inter-valley scattering is completely ignored. This assumption is valid only for Al$_x$Ga$_{1-x}$As alloys with higher Aluminium composition (x > 0.5). This is because, near $\Gamma - X$ crossover (x = 0.41-0.5), a significant amount of $X$ to $\Gamma$ back scattering takes place due to the nature of *E-k* band structure. As a result, the spin dynamics will be much more complicated and the current expression for DCP will not be valid.

Following similar approach, one can derive an expression for DCP of the photo-luminescence when the photo excitation is tuned to higher energy levels of QW. The corresponding expression is specified as,

$$DCP^{QW} = \frac{n_S^{QW}}{n_{Total}^{QW}} = \frac{P_0^{'}}{\left(1 + \frac{\gamma_S^{pure}}{\gamma_R^{QW}}\right)} \tag{10}$$

Notably, the electron spin dynamics for photo excitation inside the QW remains invariant against the nature of the barrier bandgap, represented by Fig. 1. (c). Therefore, the DCP of luminescence for photo excitation inside the QW, will be evaluated from Eq. (10) for both the cases. Further, for this photo excitation range, the value of instantaneous degree of spin polarization ($P_0'$) of photo generated electrons will be different due to modified selection rules in the QW.[26] The contribution coming from QW can be eliminated by taking the ratio of Eq. (5) and Eq. (10).

$$\frac{DCP}{DCP^{QW}} = \left(\frac{P_0}{P_0'}\right) \frac{1}{\left(1 + \frac{\gamma_S^B}{\gamma_C}\right)} \tag{11}$$

The QW contributions can differ for direct and indirect barrier samples because of (i) difference in the spin relaxation rate in QW layers, (ii) difference in compensation of unpolarized background electrons and (iii) difference in electron recombination rate in the QW. As will be shown later, all these factors contribute only a little, which is further reduced after taking a ratio as defined by Eq. (11). Once the QW contributions are carefully separated out, one can estimate the electron spin relaxation time in the barrier ($\tau_S^B = 1/\gamma_S^B$) by putting the numerical values of $DCP, DCP^{QW}, P_0, P_0'$ and $\gamma_C$ parameters. Among these, generally $DCP, DCP^{QW}$ and $\gamma_C$ are measured experimentally, whereas $P_0$ and $P_0'$ are known from the selection rules.

## III. Methods

Due to compensation of unpolarized background electrons without inducing additional ionized impurity scattering, p-type modulation doping is an attractive technique to enhance the magnitude of DCP of luminescence signal arising from QWs.[23] Hence, the samples under investigation are p-type modulation doped GaAs/Al$_x$Ga$_{1-x}$As single QWs with QW width ($d_{QW}$) of 15 nm. Aluminium (Al) composition of barrier layer is kept 0.26, 0.37, and 0.63 for the three samples labelled as S1, S2 and S3 respectively. Further, a GaAs/Al$_x$Ga$_{1-x}$As multi quantum well sample marked as S4 is also included in the investigation. This sample contains four QWs of thicknesses 2.3, 4, 8.3 and 15 nm and have Al composition of 0.63 in the barrier layer. All the samples are grown on $n^+$ -GaAs substrate (001) by MOVPE, where Trimethyl Gallium, Trimethyl Aluminum and Arsine are used as precursors to grow epitaxial layers. To achieve p type doping, Trimethyl Zinc is used as the dopant. Further, a 10 nm GaAs cap layer is grown on top of each sample to prevent the oxidization of AlGaAs barrier layer. In the sample design, spacer layer of 25 nm thickness is incorporated to separate the Zn dopants from QW. The layer is reasonably thin for the accumulation of holes in QW, but large enough for the suppression of ionised impurity scattering.[30] The numerical value of dopant density inside the QW for S3 is determined by Capacitance-Voltage method and is estimated to be $1.65 \times 10^{12}$ cm$^{-2}$ at 300K. The dopant density in other samples is kept in the similar range. QW width and Al composition of these samples are evaluated by High resolution X-ray Diffraction (HRXRD) and photo luminescence excitation (PLE) spectroscopy techniques. Pendellösung fringes observed in the HRXRD pattern, as shown in Fig. 2, indicate about a high crystalline quality of the samples. The samples are also characterized using PL technique, where only electron-heavy hole excitonic peak is observed at 10 K. The absence of electron-light hole excitonic peak indicates that the hole accumulation is minimal, but is sufficient to compensate the available free background electrons. From the width of excitonic peaks, interface fluctuation of few monolayers is estimated in all the samples.[31] Note that the usage of QW architecture of high crystalline quality significantly enhances the signal strength, thereby enabling measurement of a few percent DCP even at very low excitation intensity which is generally obtained from broad band sources like Xenon arc lamp. In our case, an average power density of only 4 mW/cm$^2$ at 630 nm is used in the measurements. The samples are mounted on cold head of a Helium based closed cycle refrigerator which can operate in 10-300 K temperature range. The details of the experimental set up are published elsewhere.[23] Here, the DCP measured PL signal is estimated by the formula $DCP = F\frac{(I^+ - I^-)}{(I^+ + I^-)}$, where, $I^+$ ($I^-$) are the PL intensity related to co (counter) polarized excitation. $F$ is the calibration factor which takes care of the misalignment of optics, frequency response of the system and its value is estimated to be 3.3 for our current experimental set up.

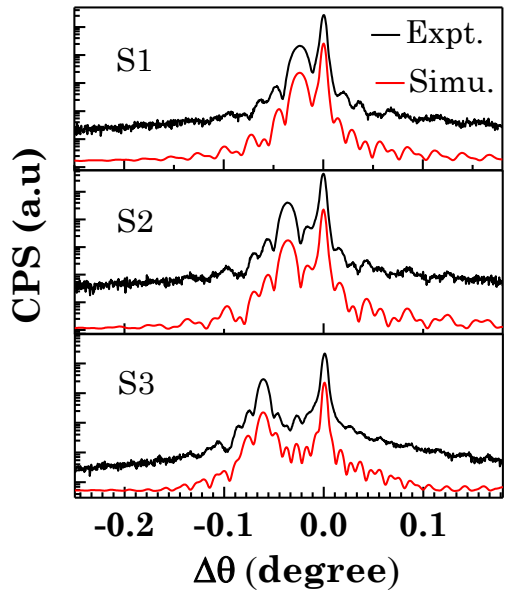

*Fig. 2*. *HRXRD pattern of GaAs/Al$_x$Ga$_{1-x}$As single quantum well samples.*

## IV. Results and Analysis

DCP of PL signal is measured for S1 and S3 samples at different values of excitation energy to investigate the electron relaxation properties. Representative 10 K DCP spectra are shown in Fig. 3 (a) and (c) for photo-excitation in the barrier and in Fig. 3 (b) and (d) for the excitation in QW layer. It can be seen from Fig. 3 (b) and (d) that in case of photo excitation inside the QW, the shape of DCP spectra for two samples remains the same apart from a shift in the energy scale due to change in the confinement potential. Peaks associated with various QW transitions are marked in the figure, which match with those observed in photo-reflectance spectra (not shown here) and are also found to be in agreement with the Eigen values obtained by solving Schrödinger equation. A similar magnitude of DCP for both the samples indicates the robustness of electron spin dynamics for photo excitation inside the QW irrespective of the difference in the nature (value) of band structure (bandgap) of the barrier layer. On the other hand, by comparing the DCP spectra of S1 and S3 for barrier excitation, one can note two important differences, (i) for near band edge excitation, magnitude of DCP in S3 is reduced by an order as compared to that of S1, (ii) Sign of the DCP spectra is opposite in S1 and S3 beyond excitation energy $E_{ex} = E_g + \Delta_{SO}$. Here, $E_g$ and $\Delta_{SO}$ stand for the band gap and split off gap near zone centre ($k = 0$) respectively. Accordingly, the impact of electron spin dynamics on the DCP spectra in these two regimes are explained separately. The results from the first regime are used to estimate *X-valley* electron spin relaxation time ($\tau_S^X$) and its temperature dependence. Theoretical calculations are also performed to corroborate with the experiments. On the other hand, the results from the second regime demonstrate the impact of *X-valley* on the hot electron spin dynamics.

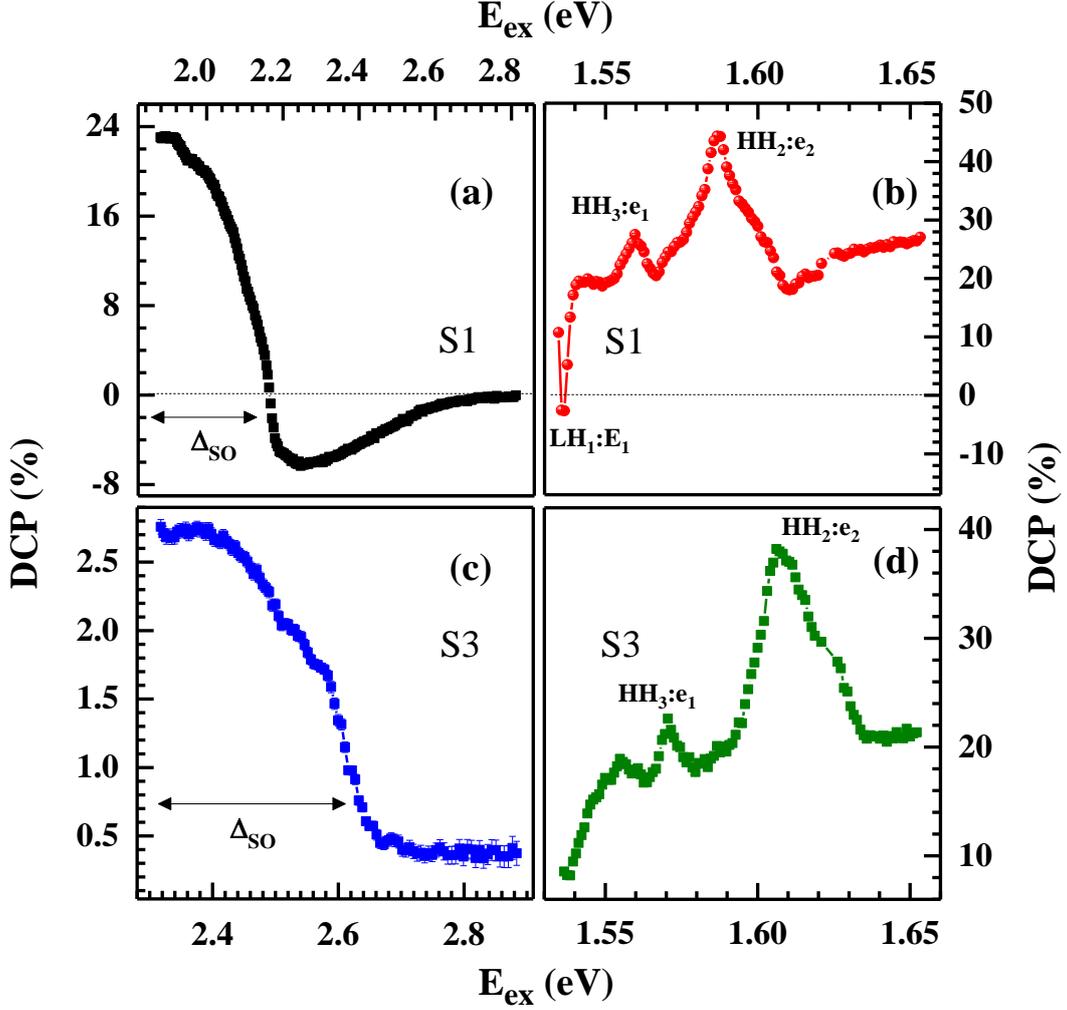

*Fig. 3*. 10 K DCP Spectra of S1 sample for (a) excitation in the barrier and (b) excitation in the QW, and of S3 sample for (c) excitation in the barrier and (d) excitation in the QW.

### (a) Estimation of electron spin relaxation time in *X-valley* ($\tau_S^X$)

For near band edge photo-excitation, the DCP spectra is determined by electrons with low kinetic energy, therefore, energy relaxation in the barrier material need not be considered. Hence, the discussion mainly involves the electron capture path from barrier to QW and simultaneous spin relaxation at various energy states inside QW. However, as explained earlier, the latter component is quite similar in both the samples. In this context, a mismatch in DCP magnitude for S1 and S3, as shown in Fig. 3. (a) and (c), can be ascribed to one of the following reasons; (i) a difference in the spin relaxation rate in the barrier layers, and (ii) a difference in the carrier capture rate from barrier to QW. The two factors are represented by $\gamma_S^B$ and $\gamma_C$ parameters respectively in Eq. (11). In literature, capture of electrons through both $\Gamma$ and $X$ valleys is studied in detail for AlGaAs/GaAs QW and the reported value of capture time $\tau_C$ ($\tau_C = 1/\gamma_C$) is about 20 ps.[32, 33, 34] Thus, from the measurement of $DCP$ and $DCP^{QW}$ and by using Eq. (11), the value of electron spin relaxation time ($\tau_S^B = 1/\gamma_S^B$) in AlGaAs barrier can be estimated. Following the proposed method, the value of $\tau_S^B$ is measured for S1, which contains AlGaAs barrier of direct band gap ($\Gamma$ valley) nature. Note that the procedure requires the values of $P_0$ and $P_0'$ also. Since $P_0$ deals with bulk barrier material, its value is taken to be 50%, which is decided by angular momentum selection rules. On the other hand, Pfalz et. al[26] have theoretically calculated the magnitudes of $P_0'$ for photo excited electrons associated with different transitions in GaAs QW. According to their calculations, $P_0' = 82.5\%$ for $HH_2$-$e_2$ transition in QW with $d_{QW} = 15$ nm. For S1, this transition is observed at $E_{ex} = 1.587$ eV in Fig. 3. (b), and the value of $DCP^{QW} = 44.4$ %. Similarly, following near barrier band edge photoexcitation in Fig. 3. (a), the value of $DCP$ is about 23%. Thus, by putting these values in Eq. (11) and taking $\tau_C = 20\ ps$, one can obtain $\tau_S^B = 1/\gamma_S^B = 118 \pm 1\ ps$. It is in good agreement with the values of $\tau_S^B$ estimated by other researchers for AlGaAs through time resolved PL and also with the values reported for GaAs whose band structure is similar to that of direct band gap AlGaAs. [22, 35] To further cross check the validity of proposed method for the measurement of $\tau_S^B$, we have selected $HH_3$: $e_1$ transition in the QW ($P_0' = 51.6\ \%, DCP^{QW} = 27.5\ \%$) and obtained $\tau_S^B = 126 \pm 1\ ps$. Thus, any higher transition showing a well-defined peak in DCP spectra can be utilized for this purpose. However, Pfalz et. al[26] have stated that thick QWs are preferred for this purpose, since in that case the spin relaxation during the thermalization of electrons via the excited states of QW is minimal.

We have followed a similar method to determine the electron spin relaxation time in the barrier layer of S3. Note that the electron spin dynamics will be governed by the satellite *X-valley* due to the indirect nature of the barrier band gap. The DCP spectra for photo excitation in the barrier and QW layers are illustrated in Fig. 3. (c) and (d) respectively. In this case, though the magnitude of $P_0$ remains 50 %, the value of $DCP^X$ turns out to be only 2.76%. Following the same procedure and considering $HH_2$-$e_2$ transition, where $P_0' = 82.5\ \%, DCP^{QW} = 38.2\ \%$ and $\tau_C^X = 1/\gamma_C^X = 20\ ps$, the estimated value of $\tau_S^X = \frac{1}{\gamma_S^X}$ turns out to be $2.7 \pm 0.1$ ps. If we consider $HH_3$-$e_1$ transition with $P_0' = 51.6\ \%$ and $DCP^{QW} = 22.7\ \%$, in place of $HH_2$-$e_2$ transition, then the value of $\tau_S^X$ turns out to be $2.9 \pm 0.1\ ps$. This further validates the proposed method. To the best of our knowledge, this is the first time that the value of $\tau_S^X$ has been measured for III-V semiconductors. It is surprising to note that even theoretical predictions for $\tau_S^X$ are not available in literature. Here, the same is tried out by considering

Dyakonov-Perel (D-P) spin relaxation under *k*-linear Dresselhaus spin splitting regime[36] and is discussed in the next sub-section of this article.

**(b) Theoretical framework for the estimation of $\tau_S^X$**

In ref. 36, it is already described that the Dyakonov-Perel spin relaxation time under *k*-linear Dresselhaus spin splitting can be calculated by using the following relation;

$$\tau_S = \frac{\hbar^4}{8m_t \beta_D^2 k_B T \tau_p} \tag{12}$$

where, $\hbar, m_t, \beta_D, k_B, T$ and $\tau_P$ stand for reduced Planck's constant, transverse effective mass of electrons, Dresselhaus spin-orbit coefficient, Boltzmann's constant, temperature, and momentum relaxation time respectively. Since, *X-valley* of AlGaAs also comprises of *k*-linear Dresselhaus spin splitting,[37] this formula could be used to estimate the numerical value of $\tau_S^X$ at 10 K. A typical value of $\tau_P = 370\ fs$ for the near band edge excitation is theoretically estimated at 10 K for *X-valley* of AlGaAs considering polar optical phonon scattering, alloy scattering, equivalent inter-X-valley scattering, deformation potential scattering, piezo-electric scattering and ionised impurity scattering in the calculations. One can calculate the *X-valley* electron mobility ($\mu^X$) using the expression $\mu^X = \frac{e\tau_P}{m_e^X}$ in order to crosscheck the value of $\tau_P$. With $m_e^X = 0.278 m_0$, the low temperature value of $\mu^X$ turns out to be $0.23\ m^2 V^{-1} s^{-1}$, which is very close to the one reported for a similar composition of AlGaAs.[38] Further, to estimate the numerical value of $\beta_D^X$, Density Functional Theory (DFT) based electronic structure calculations have been carried out using plane wave basis.[39] We have taken an energy cut off ($E_{cut}$) of 500 eV. The exchange correlational functional has been approximated by the Perdew-Burke-Ernzerhof's version of generalized gradient approximation (GGA).[40] For the sampling of Brillouin zone, we have used a Monkhorst-Pack mesh of 5×5×5 k-points. The convergence criterion for energy in the self-consistent-field cycle is taken to be $10^{-6}$ eV. To simulate the material close to the experimental stoichiometry, we have used a supercell where the required number of Al atoms have been replaced by Ga atoms. Various possible configurations have been taken into account to find out the lowest energy structure and the band structure has been calculated for that energetically lowest configuration. The band structure is calculated using a set of discrete *k* points around the conduction band minimum (CBM). The band splitting energy for the n$^{th}$ band with a momentum *k* is estimated by taking the energy difference between the bands split due to the spin-orbit coupling. The *X-valley* Dresselhaus spin-orbit coefficient is calculated by using the formula,[18] $\beta_D^X = \frac{\Delta E(k)}{2k}$, where $\Delta E$ is the spin-orbit splitting near CBM and $k$ is the lattice wave vector. The variation of $\beta_D^X$ against the wave vector $k$ is illustrated in

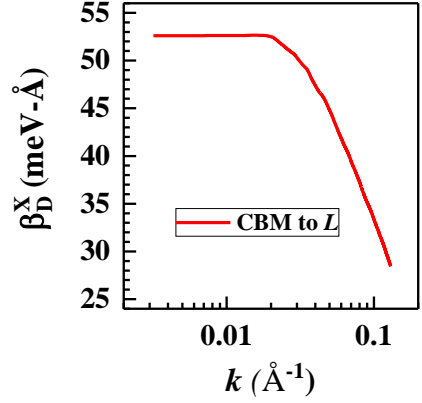

**Fig. 4**. *X-valley Dresselhaus spin-orbit coefficient along CBM to L is plotted as a function of wave vector k.*

Fig. 4. Note that the $k = 0$ point in the plot is set as the CBM where $\beta_D^X$ is found to be constant at $52\ meV$-Å before a gradual fall at higher $k$ values. This variation is similar to the one observed in GaAs.[18] The CBM value of $\beta_D^X$ is used in the calculations, since in steady state, the electrons thermalize to CBM. Other parameters like $m_t^x = 0.205\ m_0$ ($m_0$- free electron mass), $T = 10$ K, $k_B = 1.38 \times 10^{-23} J/K$ are used to get 10 K numerical value of $\tau_S^X$ to be 2.3 ps, which is found to be in good agreement with the values measured in previous section. It is worth mentioning that Eq. (12) is derived for bulk material without strain, where only Dresselhaus splitting is considered. However, presence of strain brings additional spin orbit term, which is also $k$-linear in nature.[41] In present sample, $Al_{0.63}Ga_{0.37}As$ layer is pseudomorphically grown on GaAs substrate and a residual strain of only 0.08% is measured. It might alter the estimated value of $\tau_S^X$, but the difference is expected to be rather small and the same is therefore neglected in theoretical calculations. Further, the DFT calculations are carried out by assuming an ideal system at zero Kelvin temperature, which can lead to some difference between the theory and experiments. Furthermore, a better match can be obtained by considering a slightly lower value of $\tau_C$, which is possible for the samples grown under different conditions.

**(c) Temperature dependence of $\tau_S^X$**

The temperature dependence of $\tau_S^X$ can be studied by analysing the behaviour of DCP spectra as a function of temperature. Prior to do that, the temperature sensitivity of $\tau_C$ should be clearly addressed. The temperature variation of $\tau_C$ depends upon the nature of electron capture, which can be either classical i.e. a process governed by the drift and diffusion of electrons or quantum mechanical i.e. a process governed by resonant quantum capture. One possible way to figure out the dominant mechanism is to look for the dependence of $\tau_C$ on width ($d_{QW}$) of QW. Notably, for classical electron capture, the value of $\tau_C$ does not depend upon the $d_{QW}$ unlike the quantum capture process where a clear oscillation against $d_{QW}$ is generally observed. [33] To identify the governing mechanism, we have grown a multi quantum well sample, where QWs with four values of thickness i.e. $d_{QW}$ = 2.3, 4, 8.3, and 15 nm are grown in a sample labelled as S4. Measurements and analysis are repeated for S4 similar to single QW samples. Since barrier material is same for all the QWs, $\tau_S^B$ can be taken to be the same for all of them. Further, using Eq. (11), ratio $\frac{\tau_C}{\tau_C^{15}}$ is estimated and is plotted in Fig. 5. Note that the parameter $\tau_C^{15}$ stands for the electron capture time in 15 nm thick QW. A near unity value of $\frac{\tau_C}{\tau_C^{15}}$, irrespective of different values of $d_{QW}$, indicates towards the classical capture process of electrons in this sample. It was already shown that for thick barriers like in our sample, electrons are captured through classical process which is robust against temperature over 10-300 K range. [42, 43] Thus, no temperature variation is anticipated for $\tau_C$. Since the barrier structure of S3 is identical to S4, electron capture dynamics in S3 is also expected to be similar. Therefore, by keeping $\tau_C$ = 17 ps for comparison purpose, temperature dependence of $\tau_S^X$ is studied over 10 – 80 K range and the same is plotted in Fig. 6, where a monotonous fall of $\tau_S^X$ is observed. Shortening of electron spin relaxation time with temperature has been reported earlier and the same is explained by considering the enhancement of thermal velocity of electrons.[13] A large enhancement in the thermal velocity of electrons is responsible for the rise of effective magnetic field arising from Dresselhaus spin-orbit coupling. Due to this factor, a fall in the values of electron spin relaxation time is noticed with temperature. This factor is represented by the temperature term in the denominator of Eq. (12). However, the power law for the temperature dependence of $\tau_S^X$ i.e. of $\tau_S^X \propto T^\alpha$ is primarily governed by the variation of $\tau_P$ with temperature. For the present case, the estimated value of $\alpha$ is $\sim -0.4$, indicating that $\tau_P$ varies as $T^{-0.6}$. Interestingly, no scattering mechanism support such a temperature dependence of $\tau_P$. Only alloy scattering, piezo electric scattering and inter-X-valley scattering mechanisms come close to such a trend where $\tau_P \propto T^{-0.5}$.[44] Among these scattering mechanisms, the inter-X-valley scattering is calculated to be the fastest one and exhibits a numerical value of 700 fs at 10 K. Similar conclusions are already available in literature for GaAs.[45] Hence, it is expected that the main contribution of $\tau_P$ usually comes from equivalent inter-X-valley scattering. A minor difference observed between the experimental and theoretical values of $\alpha$ can be explained by considering the effect of residual strain in AlGaAs layers due to the lattice mismatch with GaAs substrate. Kesteren et al. [46] have shown that hetero-epitaxial strain induced band structure modification in AlAs layers, keeping growth direction along the z axis, lifts the degeneracy in *X-valley* in such a way that $X_z$ lies above $X_x$ and $X_y$ on the energy scale. For $Al_{0.63}Ga_{0.37}As$, the energy separation between $X_z$ and $X_x/X_y$ is approximately 10 meV. Thus, at very low temperature (10 - 40 K), the equivalent inter-X-valley scattering will take place between $X_x$ and $X_y$ only. However, with the rise of temperature, role of $X_z$ in the inter-X-valley scattering will also become important. In this context, inter-X-valley scattering time formula can be written as,[44]

$$\frac{1}{\tau_{XX}} = \frac{Z_f D_{XX}^2 (m_e^*)^{3/2} (k_B T)^{1/2}}{\pi \sqrt{2} \rho \omega_{LO} \hbar^3} [N + (N+1)] \quad (13)$$

where, $Z_f, D_{XX}, m_e^*, N, \rho, \omega_{LO}$ stand for number of final valleys available for scattering, *X-valley* deformation potential field, effective mass of electrons in X valley, phonon number, mass density of AlGaAs and Longitudinal optical phonon angular frequency respectively. In the present scenario, $Z_f$ itself depends on temperature and its variation can be understood by the expression,

$$Z_f = \left(1 + e^{-\Delta E_X / k_B T}\right) \quad (14)$$

where, $\Delta E_X$ is the energy separation between $X_z$ and $X_x/X_y$. It can be seen that at high temperature, the value of $Z_f$ increases significantly and it makes the temperature variation of $\tau_P$ faster than anticipated. This is the reason why temperature dependence of $\tau_S^X$ in Fig. 6 deviates from $\alpha = 0.5$, especially when the temperature is above 50K.

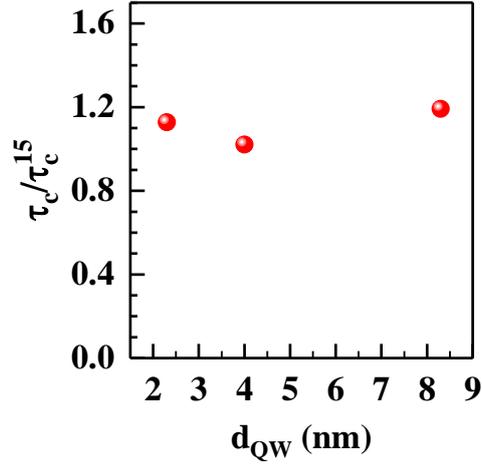

*Fig. 5. Variation of estimated electron capture time in S4 plotted as a function of QW width*

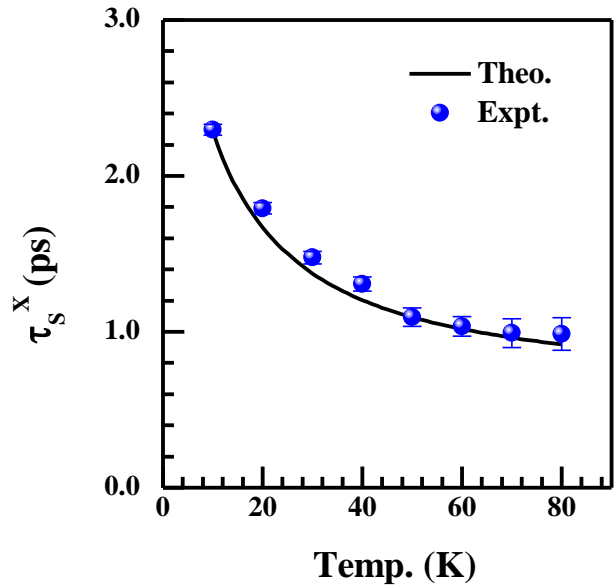

*Fig. 6. Temperature dependence of $\tau_S^X$ in S3 sample. Black solid line addresses the same for theoretically calculated values.*

**(d) Impact on hot electron spin dynamics**

From numerous studies of hot electron spin dynamics in $\Gamma$ valley, it is clear that the D.P spin relaxation enhances with the kinetic energy of electrons.[25] One can conveniently inject hot electrons in the conduction band by keeping the energy of the impinging photons much larger than the bandgap of semiconductors. Consequently, the DCP spectra generally falls with excitation energy and the nature of spectra is expected to depend on the spin splitting and energy relaxation time. Since for *X* valley, these two parameters are significantly different from the $\Gamma$ valley, the corresponding DCP spectra are expected to be different particularly at high excitation energy. One of the key differences between the two cases is seen for $E_{ex} = E_g + \Delta_{SO}$, where the observed 10 K DCP spectra of S1 and S3 is of opposite sign, as shown in Fig. 3 (a) and Fig. 3 (c). Since the QW material and thickness are same in the two samples, the difference is bound to be from the barrier material alone. Ekimov et al.[8] have pointed out that in GaAs, the sign of the DCP spectra above $E_{ex} = E_g+\Delta_{SO}$ depends upon the dopant density of the sample and it changes beyond a critical value of dopant density. However, in the present scenario, AlGaAs barriers are un-intentionally doped (<$10^{15}$ cm$^{-3}$), except for the thin doped spacer layers. Thus, the different nature (direct versus indirect) of band gap of AlGaAs barriers in S1 and S3 is ought to be the fundamental reason behind this observation since the electrons are captured via *X-valley* in the latter case. To further crosscheck this observation, similar measurements are performed in S2 which bears an AlGaAs barrier of Al composition 0.37. For this Al composition, AlGaAs barrier is direct band gap but very close to *Γ-X* crossover.[47] Measured DCP spectra for this sample is shown in Fig. 7 and a clear negative value is observed beyond $E_{ex} = E_g+\Delta_{SO}$.

Note that the optical orientation properties of electrons near *Γ* valley are well known and will be helpful in understanding the physics behind these results. Here, the fall of DCP with excitation energy can be explained by considering the transitions from various valence bands having different angular momentum. For, near band edge ($k = 0$) excitation energy, electrons are excited from heavy hole (*HH*) and light hole (*LH*) bands. At Brillouin Zone centre ($k = 0$), both these bands are equal weightage admixture of $\left|\frac{1}{2},\frac{1}{2}\right\rangle$ and $\left|\frac{1}{2},\frac{-1}{2}\right\rangle$ angular momentum states.[10] Calculations considering the selection rules lead to the fact that for near band edge, both the *HH* and *LH* bands will generate electrons with 50% spin polarization. However, with increase in photon energy (i.e. for transition involving larger *k*) the *LH* band acquires more and more split off (*SO*) band character and therefore generates electrons with opposite spin polarization.[10] This will cause a decrease in the degree of spin polarization of photo generated electrons and the same is reflected in the DCP spectra shown in Fig. 3 and Fig. 7.

To explain the experimental results beyond $E_{ex} = E_g+\Delta_{SO}$, we recall Eq. (5) which estimates the value of DCP of luminescence for near band edge excitation but excludes the electron spin loss during thermalization process. However, for hot electrons as in the present case, this process becomes much more important and these equations will lead to erroneous values of DCP. Thus, to invoke the physics of electron spin relaxation during their energy relaxation, Eq. (5) is to be modified as given below, [23]

$$DCP = \frac{P_0 \exp\left[-\int_{E_{th}}^{E_0} \frac{\gamma_E^B(E_k)}{\gamma_S^B(E_k)} \frac{dE_k}{E_k}\right]}{\left(1 + \frac{\gamma_S^{QW}}{\gamma_R^{QW}}\right)\left(1 + \frac{\gamma_S^B}{\gamma_C}\right)} \qquad (15)$$

where, $\gamma_E^B$, and $E_k$ are the energy relaxation rate of electrons in the AlGaAs barrier, and kinetic energy of photo generated electrons. $E_0$ and $E_{th}$ stand for the kinetic energies of electrons at the time of photo excitation and after reaching equilibrium. This expression is valid for both direct and indirect band gap AlGaAs barrier. The only difference is the region of operation of electrons which is Γ valley for direct band gap AlGaAs and *X-valley* for indirect band gap AlGaAs. Further, this expression clarifies that the nature and the magnitude of $DCP^\Gamma$ and $DCP^X$ beyond $E_{ex} = E_g + \Delta_{SO}$ is a signature of kinetic energy ($E_K$) dependence of electron spin relaxation time, which is cubic for *Γ* valley and linear for *X* valley. [23,37] To appreciate this fact, 14 band *k.p* calculations are performed to estimate the band structure of AlGaAs. This computation reveals that for photo excitation near $E_g + \Delta_{SO}$, the approximate kinetic energies of electrons generated from *HH*, *LH* and *SO* bands are 150 meV, 95 meV and 15 meV respectively. In case of direct band gap AlGaAs barrier (S1, S2), electrons are captured in the QW through *Γ* valley. As a result, two different bunches of photo generated electrons are observed, (i) electrons generated from *HH* and *LH* band with very high kinetic energy (ii) electrons generated from *SO* band near band edge. In the former case, most of the electrons will lose their spin orientation even before reaching the QW. On the contrary, electrons in the latter case will remember their spin polarization and thus the DCP of luminescence will reflect negative degree of spin polarization character of *SO* band. This result is a consequence of $k^3$ dependence of Dresselhaus spin splitting in $\Gamma$ valley and more details about it are published elsewhere.[23]

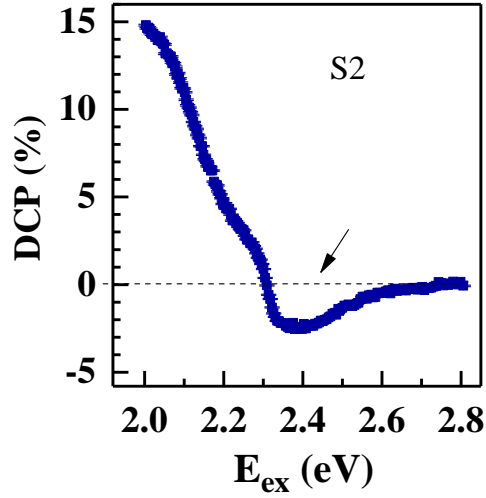

*Fig. 7*. 10 K DCP spectra of S2 sample

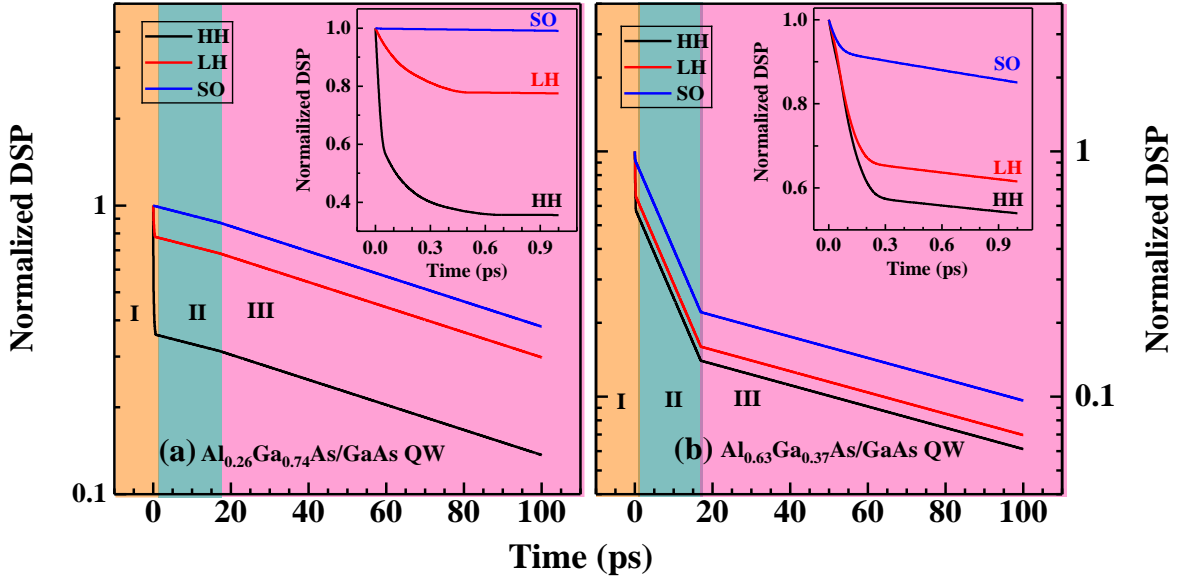

*Fig. 8. Numerically calculated temporal evaluation of electron's degree of spin polarization (DSP) in (a) S1 and (b) S3, the respective insets show the magnified portion of initial data to clarify the effect of simultaneous energy and spin relaxation, The regions shaded as orange (I), indigo (II), and pink (III), represent different regimes of electron spin relaxation process.*

This can be better appreciated if degree of spin polarization (DSP) is plotted as time evolution of spin polarization of electrons excited from different bands. This is depicted in Fig. 8. (a) and (b), which are divided into three parts, in accordance with the dynamics of the corresponding time regime. The regimes labelled as I (orange), II (indigo) and III (pink) are related to electron spin relaxation during thermalization, after being thermalized to barrier band edge and in QW respectively. Initial part of the data is magnified and plotted as insets in the respective figure. The time axis is taken up to 100 ps, which is generally the carrier recombination time in QW. Therefore, the normalized DSP at 100 ps is the DCP, which is observed experimentally. In S1 [Fig. 8. (a)], it can be seen that the electrons generated from *SO* and *LH* valence bands can retain their spin polarization significantly, whereas those excited from HH bands lose it very quickly. Therefore, the observed DCP has negative polarity due to a dominant component of DSP appearing from the *SO* band.

On the other hand, in S3, which consists of an indirect band gap AlGaAs barrier, the initial electron spin dynamics takes place in the satellite *X-valley* where the energy relaxation process is faster as compared to that of $\Gamma$ valley.[48] Thus, electrons can be readily thermalized to *X-valley* minimum irrespective of their kinetic energy and the duration, during which electrons simultaneously lose their spin polarization and energy, becomes shorter as shown in Fig. 8. (b) and its inset. Further, due to linear $k$ dependent Dresselhaus spin splitting in the *X* valley, $\tau_S^X$ is less sensitive to $E_K$. Even though the numerical value of DSP of electrons generated from *HH* at 100 ps is slightly lower as opposed to that from *LH* and *SO* electrons, the larger value of photo absorption in the former case maintains the positive polarity in the DCP spectra beyond $E_{ex} = E_g + \Delta_{SO}$. Therefore, the opposite polarity of DCP spectra at these excitation energies is directly correlated to the different nature of spin dynamics in $\Gamma$ and *X* valley.

## V.    Summary and Conclusion

In conclusion, spin relaxation time ($\tau_S^X$) in the *X-valley* of indirect bandgap AlGaAs barrier layers is measured by analysing the DCP of PL signal of QW layer, where the contribution from spin relaxation in QW is carefully eliminated by a selective photoexcitation within the QW layer. Assuming a two-level model, an analytical expression relating the DCP of PL signal and $\tau_S^X$ of barrier layer is derived. Using this equation, $\tau_S^X$ is estimated to be 2.7 ± 0.1 ps, which is found to be in good agreement with the theoretical calculations based on DFT. Temperature dependence of $\tau_S^X$ is recorded in 10 – 80 K range, which is explained by invoking D-P spin relaxation under linear spin splitting regime. It is found that intra-valley scattering in strain modified *X-valleys* lying along the in-plane and transverse directions dominates the spin relaxation process. Further, spin dynamics of photo-generated hot electrons is investigated by measuring the DCP of PL signal arising from an adjacent QW layer for both direct and indirect bandgap AlGaAs epitaxial layers. The distinct nature of DCP curve beyond $E_{ex} = E_g + \Delta_{SO}$ for the two cases is explained by considering faster energy relaxation and linear spin splitting in satellite *X-valley*. Indirect band nature of $Al_{0.63}Ga_{0.37}As$, where accumulation of spin polarized carriers is feasible in *X-valley* even in the absence of external bias, provides an excellent opportunity for the development of spin devices with adequate Joule heating. Enhanced value of $\tau_S^X$, which is an order of magnitude higher than $\tau_S^\Gamma$, is expected to help in the optical manipulation of spin devices. The two factors mentioned above can lead to a new horizon for the development of next-generation spin-photonic devices. Furthermore, the procedure elaborated here will be applicable in case of other group III-V and group IV indirect bandgap semiconductor and is expected to be useful in understanding the role of phonon-assisted processes in spin polarization of electrons.

**Acknowledgements:**

The authors acknowledge Dr. V. K. Dixit and Ms. Geetanjali for their help in sample growth and useful discussions. Technical support received from Mr. U. K Ghosh, Mr. Sanjay Porwal and Mr. Alexandar Khakha is also acknowledged. Authors further acknowledge the contribution of Dr. D. Pandey in the DFT calculations and thank the Computer Division, RRCAT for support in installing and smooth running of the code. P.M. and J.B. thank HBNI-RRCAT for financial support.